\documentclass[preprint2]{aastex}
\pdfoutput=1

\usepackage{array}
\usepackage{amsmath}

\begin{document}

\title{Early-Time Stability of Decelerating Shocks}

\author{F. W. Doss and R. P. Drake}
\affil{Department of Atmospheric, Oceanic, and Space Sciences, University of Michigan, Ann Arbor, MI 48105}
\author{H. F. Robey}
\affil{Lawrence Livermore National Laboratory, Livermore, California 94550}

\begin{abstract}
We consider the decelerating shock instability of Vishniac for a finite layer of constant density.  This serves both to clarify which aspects of the Vishniac instability mechanism depend on compressible effects away from the shock front and also to incorporate additional effects of finite layer thickness.  This work has implications for experiments attempting to reproduce the essential physics of astrophysical shocks, in particular their minimum necessary lateral dimensions to contain all the relevant dynamics.
\end{abstract}


\section{Introduction}

\cite{V83} outlined a theory of instabilities for a system of a decelerating shock accreting mass, modeled as a thin mass shell layer possessing no internal structure.  In \cite{V89}, the theory was expanded to include a layer of post-shock material, exponentially attenuating in density.  Other work (\cite{Bertschinger1986, KWS05}; among others) has described the perturbation of self-similar solutions for the post-shock flow.  The present work complements these investigations,
modeling the post-shock flow as a finite thickness layer of constant density and considering both compressible and incompressible post-shock states.  This allows us both to more clearly understand which mechanisms depend on the compressibility of the shocked gas and which are common to any shock system undergoing deceleration.  

Early in the lifetime of an impulsively driven shock, when the post-shock layer thickness is small compared to its compressible length scale, an exponential scale cannot be formed and the density profile may be closely approximated by a square wave, as a fluid everywhere of constant density.  In the shock's frame, upstream fluid is entering the shock with a speed $V_s$ and exiting it with a speed $U = V_s \eta$, where $\eta$ is the inverse compression ratio  associated with the shock, including both the viscous density increase and any subsequent, localized further density increase in consequence of radiative cooling \citep{DrakeHEDP}.

We have also in mind throughout this paper experiments \citep{Reighard06, BouquetPRL, BozierPRL} that have been carried out to investigate radiating shock dynamics.  Experiments of this type are often designed to be scaled to relevant astrophysical investigation \citep{RemingtonRMP}.  The particular experiments of \cite{Reighard06} feature characteristic shock velocities of over 100 km/sec, shock tubes with 625 $\mu$m diameters, and strong deceleration throughout the shocks' lifetimes.

\section{System of a Decelerating Shock with a Dense Downstream Layer}

\begin{figure}[tbp] 
   \centering
   \includegraphics[width=\columnwidth]{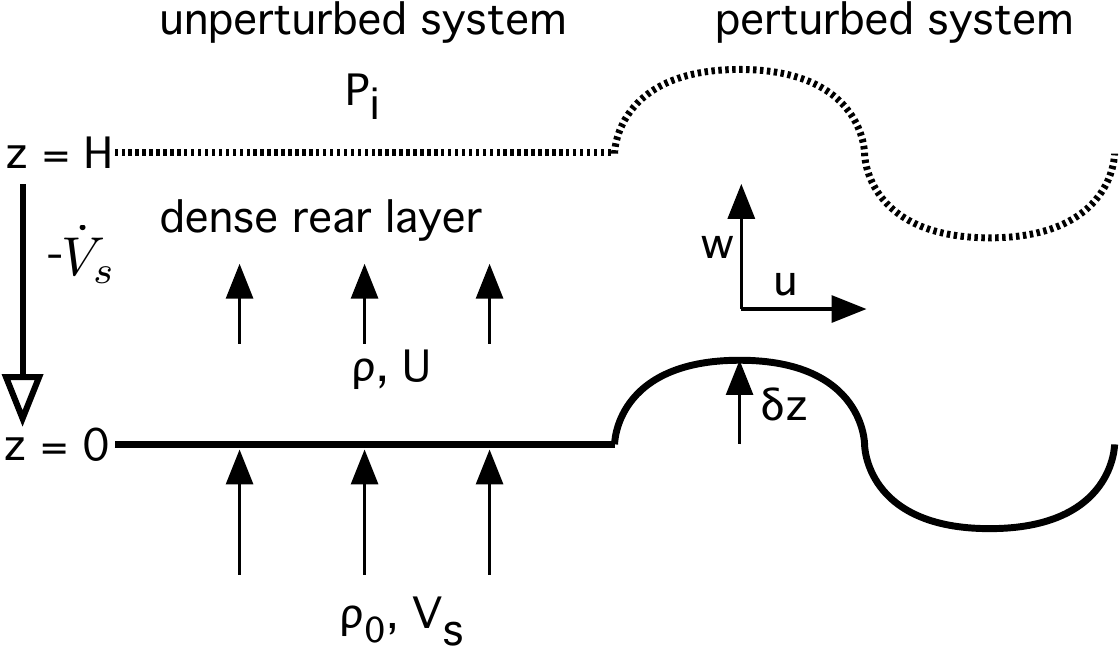} 
   \caption{Schematic of the decelerating shock system.  The solid black line is the shock, the dashed line above the dense rear layer is the rear material interface.  The left-hand arrow depicts the in-frame inertial force with acceleration $(-\dot{V}_s)$.}
   \label{system}
\end{figure}

We consider the shock in its own, decelerating frame.  The system is depicted in Figure \ref{system}.  The shock is placed at $z=0$, with flow entering it from the negative $z$ direction at speed $V_s$, density $\rho_0$, and with negligible thermal pressure.  Flow is exiting the shock toward positive $z$ with speed $U$, density $\rho$, and isotropic pressure $P$.  We will model the downstream, rear layer as a constant density region of finite, increasing thickness from $z=0$ to $z=H$.  The rear surface of the dense layer will be taken to be a free interface at constant pressure.  Beyond the rear layer will be taken as a region of constant thermal pressure $P_i$.

The native surface wave modes in the system will be right- and left-propagating waves on the two surfaces of the dense layer, leading to four modes in total.  As drawn in Figure \ref{system}, the upper surface, a material discontinuity, is stable if the shock frame is decelerating and is characterized by surface gravity modes.  The bottom surface, a shock at which compressibility is not suppressed, will feature propagating acoustic modes.  The waves which appear in our coupled system will be modifications of these waves which appear on these surfaces in isolation.  In particular, the modified acoustic waves along the shock surface will be identified as bending modes of the entire dense layer.

In order to understand the fundamental cause of the instability, we will here be considering the fluid both ahead and behind the shock to be held at (different)  densities constant in both space and time.  This practice is described and defended by \cite{Hayes}, who in their book ``consider constant-density hypersonic flows, though we should never consider the fluid in a hypersonic flow as incompressible.''  The pressure profile behind the shock is hydrostatic, $P(z) = P_i - (H - z) \rho \dot{V}_s $, which leads to increasing pressure at the shock front when the shock is decelerating.  Perturbations to density by the waves under investigation will be discussed.

\section{Linear Perturbations of the System}

\subsection{Solutions Inside the Post-Shock Fluid} \label{SolnsInsideSection}

We begin with the inviscid fluid equations
\begin{eqnarray}
\rho (\partial_t {\underline v} + {\underline v} \cdot \nabla {\underline v}) = - \nabla P - \rho {\dot V_s} {\hat z} \\
\partial_t \rho + {\underline v} \cdot \nabla \rho = - \rho \nabla \cdot {\underline v}
\end{eqnarray}
with total velocity ${\underline v} = (u,0,w + U)$ and $P = P + \delta P$.  We will insert the perturbation $\delta \rho$ only in the continuity equation; the coupling of $\delta \rho$ to the frame's acceleration will be suppressed.  This allows us to ignore mode purely internal to the layer, concentrating on the overall shock and layer system.  
Since log $\rho / \rho_0 >>$ log $ (\rho + \delta \rho) / \rho$ for any reasonable density perturbations, we expect the dynamics of the system to be dominated by the compression at the shock.
The omission of the term $\delta \rho \dot V_s$ is also required for consistency with the assumption of our square wave density profile; the system will otherwise begin to evolve into an exponential atmosphere.

We first let the perturbations $u, w, \delta P$ have time and space dependence as $e^{n t + i k x}$, with $k$ real and $n$ complex.  We then linearize the $x$- and $z$- components of the momentum equation to obtain
\begin{equation}\label{xmom}
(n + U \partial_z)u = - \frac {i k \delta P} {\rho} 
\end{equation}
\begin{equation}\label{zmom}
(n + U\partial_z) w + w \partial_z U = -  \frac {\partial_z \delta P}{\rho}
.\end{equation}
We expressed the perturbed continuity equation in terms of perturbed pressure,
\begin{equation} \label{cont}
i k u + \partial_z w = - \frac {(n + U\partial_z)\delta \rho }{\rho} = - \frac {(n + U\partial_z)\delta P} {\rho c_s^2 }
\end{equation}
where $c_s^2 = \partial P / \partial \rho$.  We solve
Equation \ref{xmom} for $\delta P$ using Equation \ref{cont}, and discard terms of order $U / c_s$ to obtain 
\begin{equation} \label{pres}
\delta P = \frac {\rho}{k^2 + n^2 / c_s^2} (n +  U\partial_z)(-\partial_z w)
\end{equation}
and insert that into
Equation \ref{zmom} to obtain a new equation for z-momentum:
\begin{equation} \label{zmom2}
(n + U \partial_z)w + w\partial_z U =  \partial_z\left( \frac {1}{k^2 + n^2 / c_s^2} (n+U\partial_z)(\partial_z w) \right)
\end{equation}
which can now be written as a differential equation for $w$ (taking $U$ and $c_s$ constant throughout the post-shock layer),
\begin{equation} \label{goveq}
\left(\frac{U}{k^2 + n^2/c_s^2}  \partial^3_z + \frac {n}{k^2 + n^2/c_s^2} \partial^2_z - U \partial_z - n\right)w = 0
.\end{equation}
For a treatment of the problem where $c_s$ varies through the layer, see Appendix A.

We define
\begin{equation}
j = \sqrt{k^2 + n^2 / c_s^2},
\end{equation}
which describes the effective lateral wavenumber.  As a wave approaches the acoustic case, $n^2 = -k^2 c_s^2$, the wave becomes purely longitudinal and $j$ tends toward zero.
Equation \ref{goveq} has the general solution 
\begin{equation} \label{solns}
w = A e^{j z} + B e^{-j z} + C e^{- {n z / U}}.
\end{equation}
The system accordingly has three boundary conditions at its two interfaces: the shock and the rear surface.  We note that the shock frame's acceleration $\dot{V}_s$ does not appear in the general form of the perturbations; it will enter into the system through the boundary conditions.

The last term in Equation \ref{solns} is a consequence of the background flow $U$ and is closely connected with structures convecting downstream with that velocity.   It is instructive to consider the general solution for $w$ in the frame of the rear surface.  We introduce the coordinate $z' = U t - z$.  In addition, we will now write explicitly the implicit time-dependence $e^{nt}$.  The general solution is $w = A e^{(n + jU) t  - j z' + n t } + B e^{ (n  - jU) t + j z'} + C e^{ {n z'}/{U} }$.  We see that the third term has no time-dependence in the frame of the rear layer.  In the frame of the rear surface, these flow structures are generated by perturbations in the shock surface as the shock passes some point in space, and do not evolve further.  Therefore, in the frame of the shock, this term describes flow structures convecting downstream through the flow with constant velocity $U$.  We take the shock to have been perfectly planar at the instant, some time past, at which the shock's deceleration and rear layer formation began.  This allows us to explicitly set $C = 0$ at the rear layer.  We assume however that the perturbation began sufficiently early in time that our treatment using Fourier modes is sufficient, so no further information from initial conditions will be incorporated at this time.

\subsection{Infinitely Thin Layer} 

We recall that the dispersion relation for the thin shell instability in its most simple form, without the effects of compression, is in \cite{V89} written in the form
\begin{equation} \label{Vdisp}
n^4 + n^2 c_s^2 k^2 -  \frac {k^2\dot{V}_s P_i }{ \sigma} = 0
\end{equation}
where $\sigma$ is the areal mass density of the (infinitely) thin layer, and all other variables are as we have defined them.  Early work \citep{V83} derived this expression for a shock of infinitesimal height but finite areal density.  Such a shock, maintaining an infinitely thin layer height while continuing to accrete mass from the incoming flow, would in our analysis be described as the limit of an infinite compression, $\eta \rightarrow 0$.   We should expect solutions we obtain for layers of finite thickness to approach Equation \ref{Vdisp} in this limit.

\subsection{Free Rear Surface} \label{ThinLayerSection}

We construct the boundary condition describing a free layer at $z = H$ by applying $\delta P = \rho (- \dot V_s)\delta z$ at $z = H$, with $\partial_t \delta z = w$.  Using Equation \ref{solns} and our earlier expression for $\delta P$, Equation \ref{pres}, the boundary condition becomes
\begin{eqnarray} \label{wavecond}
A(n^2 - j \dot{V}_s)e^{j H} + B(-n^2 - j \dot{V}_s)e^{-j H} = 0
\end{eqnarray}
where $C$ has been explicitly set to zero as discussed above.
Equation \ref{wavecond} is a boundary condition well known to generate surface gravity waves, when $j = k$ and when paired with a rigid boundary condition at $z=0$.

At the shock surface, we must perturb the shock momentum jump condition in the frame of the moving shock.  The perturbed shock surface moving upward in Figure \ref{system} sees a weaker incoming flow.  In addition, by raising the shock surface in the hydrostatic pressure field, the effective post-shock pressure drops by an amount $\dot{V}_s \rho \delta z$.  Our jump condition has now become
\begin{mathletters}
\begin{equation}
\rho_0 (V_s - w)^2 = \rho U^2 + (P  + \dot{V}_s\rho \delta z + \delta P)
,\end{equation}
from which we obtain a boundary condition (using $\rho_0 V_s = \rho U$, $\delta z = w/n(1-\eta)$, and  our earlier expression for $\delta P$ in Equation \ref{pres})
\begin{equation} \label{prescond}
\left.\left( \frac {U}  {j^2} \partial^2_z + \frac {n }{ j^2} \partial_z  -  \left( \frac {\dot{V}_s }{ n (1 - \eta)} + 2 U\right)\right)  w\right|_{z=0} = 0
.\end{equation}
\end{mathletters}
The expression for $\partial_t \delta z$ comes from conservation of mass across the shock.  With density perturbations suppressed, as discussed above, we have a balance of mass flux with $\rho_0 V_s$ entering and $\rho U + w$ leaving the shock, with the shock moving at speed $\partial_t \delta z$.
\begin{mathletters}
\begin{equation}
\eta = \frac { U} { V_s} = \frac {U + w - \partial_t \delta z }{V_s - \partial_t \delta z} \\
\end{equation}
implying (with $\partial_t = n$)
\begin{equation} \label{shocketa}
\left. w\frac{{}}{}\right|_{z=0} = (1 - \eta) n  \delta z
\end{equation}
\end{mathletters}

The third boundary condition comes from oblique shock relations.  Letting $\beta$ be the angle of the shock surface perturbation, continuity of the tangential flow requires to first order
$u \approx V_s \beta = (i k) V_s \delta z .$
Applying the continuity equation of Equation \ref{cont} just downstream of the shock,  and applying Equations \ref{pres} and \ref{shocketa}
\begin{mathletters}
\begin{eqnarray}
\left.\left(\partial_z - \frac {V_s k^2 }{ n (1-\eta)}\right) w\right|_{z=0} = -  \frac{(n + U\partial_z) \delta P} {\rho c_s^2}
\end{eqnarray}
which evaluates to
\begin{eqnarray} \label{shockcond}
\left. \left( \partial_z - \frac {V_s j^2}{n(1-\eta)} \right)\ w   \frac{{}}{}\right|_{z=0}  =  0 
\end{eqnarray}
\end{mathletters}

Simultaneously applying these three conditions (equations \ref{wavecond}, \ref{prescond}, and \ref{shockcond}) on $w$, one demands for nonzero solutions that the determinant of the matrix of coefficients of $A$, $B$, and $C$, shown collected in Equation \ref{matrix}, must be zero,
\begin{equation} \label{matrix}
\left|
\begin{array}{ccc}
(n^2 - j \dot{V}_s)e^{j H} & (-n^2 - j \dot{V}_s)e^{-j H} & 0 \\
 -\frac{n }{j} +U +\frac{{\dot V_s} }{n (1 - \eta)} & \frac{n }{j} +U  +\frac{\dot V_s}{n(1 - \eta)} & 2U + \frac{\dot V_s}{n(1 - \eta)} \\
 \frac {1} {j} - \frac{V_s}{n( 1 - \eta)} & - \frac {1} {j} -\frac{V_s}{n(1- \eta)} & -\frac{n}{ Uj^2 }-\frac{ V_s}{n(1-\eta)}
 \end{array}
\right| = 0
\end{equation}

From this one obtains, with some manipulation, the dispersion relation,
\begin{equation} 
\begin{split}
0 = (1 - \eta)n^2 + j^2 U V_s + (j\dot{V}_s + 2 n j U) \times  \\
\left( \frac 
{
 (n^3 + j^2U\dot{V}_s) - (n j \dot{V}_s + n^2jU)\tanh jH
}
{
 (n^3 + j^2U\dot{V}_s)\tanh jH - (n j \dot{V}_s + n^2jU)
} \right) \label{wavedisp}
.\end{split}\end{equation}
We will take, as in \cite{V89}, the product $U V_s$ to be equivalent to an average sound speed squared $\langle c_s^2 \rangle$, which we shall not henceforth distinguish from the sound speed $c_s^2$ of material compressibility. 
The qualitative classification of solutions to Equation \ref{wavedisp} depends strongly on the layer thickness $H$, specifically on its relation to the compressible scale height $U V_s  / |\dot{V}_s| =  c_s^2 / |\dot{V}_s|$.  We shall explore this dependence in what follows.

\begin{figure}
   \includegraphics[width=\columnwidth]{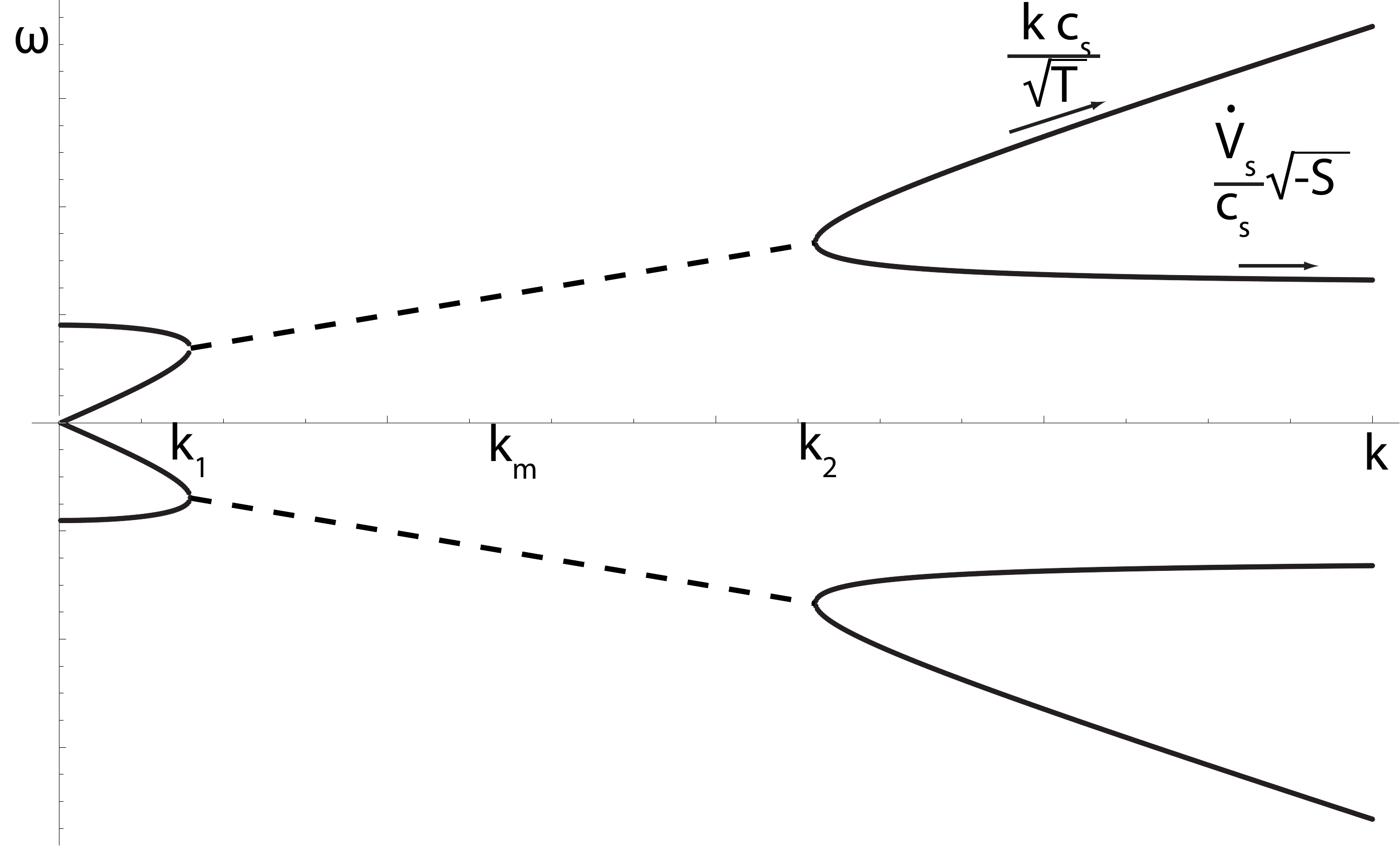} 
   \caption{Plot showing solutions of Equation \ref{cVdisp}, $\omega = \textrm{Im}(n)$ \textit{vs.} $k$.  The dashed line denotes the region of instability, where Re($n$) is nonzero.}
   \label{cVplot}
\end{figure}

We will now investigate the range in which wavelengths of perturbations are not much shorter than $H$, and will approximate $\tanh j H \approx j H$.  The existence of the critical $H$ is easiest to see in the limit of very strong, highly compressive shocks ($U \rightarrow 0$ while $V_s \rightarrow \infty$ in such a way that $U V_s =  c_s^2$ and $\dot V_s$ remain constant).
By expanding $j$, we may write the dispersion relation as,
\begin{mathletters} \label{cVdisp}
\begin{eqnarray}
T n^4 + n^2 \left( k^2 c_s^2  -  \frac {\dot V_s^2} {c_s^2} SZ \right) - k^2 \dot V_s^2 S = 0
\end{eqnarray}
where we have introduced scale factors
\begin{align}
T &= 2 - \eta \\
S &=  1 + \frac {c_s^2 / \dot V_s}{H}  \\
Z &=  1 - \frac {\eta (S-1)}{S} 
.\end{align}\end{mathletters}
For strong shocks, $T  \sim (\gamma + 3)/(\gamma + 1)$, in which any effects of strong radiation are included in $\gamma$ as an effective polytropic index describing the total density increase at the shock \citep{Liang}.  $Z$ is typically close to 1.  Solutions of Equation \ref{cVdisp}, shown in Figure \ref{cVplot}, yield instability for $k$ in the range $k_1 < k < k_2$, centered around a wavenumber of maximum instability $k_m$, where
\begin{mathletters}
\begin{align}
k_1 &= \frac {|\dot V_s| \sqrt{-S}}{c_s^2} \sqrt{2T - Z - 2\sqrt{T^2 -TZ}} \\
k_2 &= \frac {|\dot V_s| \sqrt{-S}}{c_s^2} \sqrt{2T - Z + 2\sqrt{T^2 -TZ}} \\
k_m &= \frac {|\dot V_s| \sqrt{-S}} {c_s^2} \sqrt{T} 
\end{align}
\end{mathletters}
We find that $k_1$ and $k_2$ are real for $S < 0$, requiring $\dot V_s < 0$ and $H < c_s^2 / |\dot V_s|$, conditions defining a decelerating shock and a layer width shorter than a scale height.  

For the high compression limit $T = 2$, $Z = 1$, the critical wavenumbers take the values
\begin{mathletters} \label{cVks}
\begin{align}
k_1 &= \frac {|\dot V_s| \sqrt{-S}}{c_s^2} \sqrt{3 - \sqrt{8}} \sim 0.293 k_m\\
k_2 &= \frac {|\dot V_s| \sqrt{-S}}{c_s^2} \sqrt{3 + \sqrt{8}} \sim 1.707 k_m\\
k_m &= \sqrt{2}  \frac {|\dot V_s| \sqrt{-S}} {c_s^2}
\end{align}
\end{mathletters}
The solutions for growth rate at the fastest growing wavelength are 
\begin{equation} \label{cVn}
n_m = \pm \left( \sqrt{\frac 1 8} \pm i \sqrt{\frac 7 8} \right) \frac {|\dot V_s| \sqrt{-S}}{c_s}
\end{equation}
which allow us to verify that for $k = k_m$,  $j H = (k_m^2 + n_m^2/c_s^2) H$ remains small, validating our assumption.

In the opposite limit of shock strength, as $\eta = 1$ and the shock is removed from the system, $T = 1$ and the bending waves asymptotically approach unmodified acoustic waves at high $k$.  

\subsection{Limiting Behavior of Solutions}

\begin{figure}
   \includegraphics[width=\columnwidth]{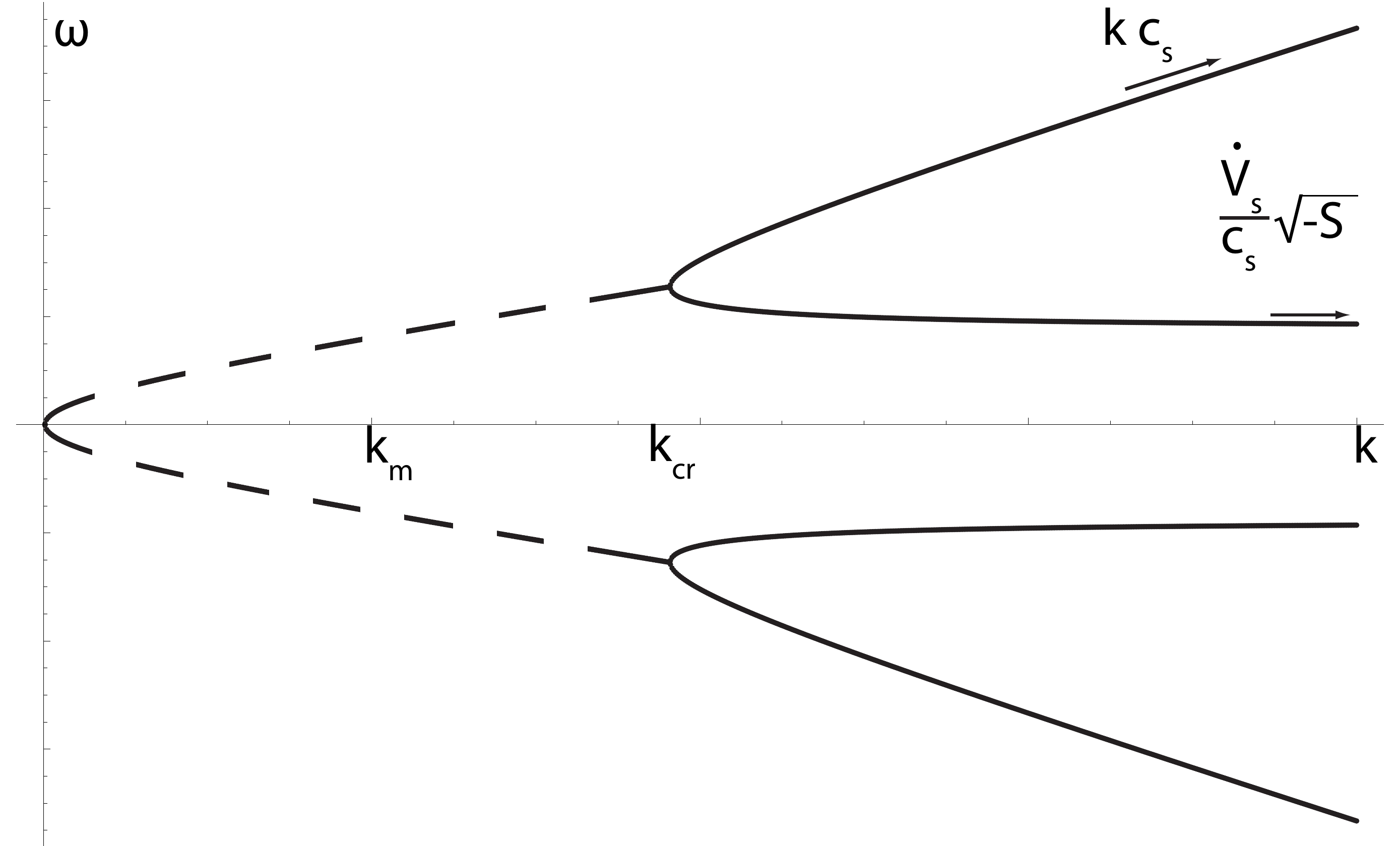} 
   \caption{Plot showing solutions of Equation \ref{criticaldisp}, $\omega = \textrm{Im}(n)$ \textit{vs.} $k$.  The dashed line denotes the region of instability, $k < k_{cr}$, where Re($n$) is nonzero.}
   \label{incVplot}
\end{figure}

To investigate individually the effect of various terms, we may make several further simplifying assumptions to Equation \ref{wavedisp}.  We will consider both the high compression limit, $\eta \rightarrow 0$, as well as the limit of negligible compressibility, $j^2 \rightarrow k^2$.  We then obtain from Equation \ref{wavedisp} the equation
\begin{equation}
\begin{split}
0 = n^2 + k^2  c_s^2  + (k\dot{V}_s + 2 n k U) \times  \\
\left( \frac 
{
 (n^3 + k^2U\dot{V}_s) - (n k \dot{V}_s + n^2Uk)\tanh kH
}
{
 (n^3 + k^2U\dot{V}_s)\tanh jH - (n k \dot{V}_s + n^2Uk)
} \right) \label{wavedisp2}
.\end{split}\end{equation}

The $2nUk$ term in Equation \ref{wavedisp2}, which stems from the same physical source as the term discarded in Equation 6b of \cite{V89}, contributes to damping and shock stability in the high $k$ limit.  It was demonstrated in early work, such as that by \cite{Freeman1955}, that we expect stability for shocks separating two simple spaces of homogenous material.  Accordingly, in systems with decelerating shock-bounded dense layers, as we tend to wavelengths short compared to the width of the layer, the dynamics must approach this stable limit \citep{V1995}.
The correct rate of damping  is however beyond the scope of our assumptions.  \cite{IshizakiPRL} have shown that the acoustic modes within the shocked material, which we have suppressed, play a role in stabilizing the shock.

The limit of an indefinitely thin layer is approached, in the notation of Equation \ref{wavedisp2}, by taking  the limit of negligible post-shock flow $U \rightarrow 0$, rearranging the dispersion relation as
\begin{eqnarray} \label{criticaldisp}
n^4 + n^2 k^2  c_s^2 - k^2 \dot{V}_s^2\left[1 + \frac { c_s^2 / \dot{V}_s} {H} \right] = 0
.\end{eqnarray}
This shows that, in these limits, we regain the form of the Vishniac dispersion relation (Equation \ref{Vdisp}).  We also see that, for $H$ less than the scale height and $\dot V_s < 0$, the quantity in square brackets becomes negative, while this quantity is positive for large $H$ or positive $\dot{V}_s$.  This means that the solutions to Equation \ref{criticaldisp} have the signature of the Vishniac thin layer dispersion for an \textit{accelerating} shock, except when $H$ lies within a scale height for a decelerating shock, $H <  -  c_s^2 / \dot{V}_s$.  Solutions when $H$ is in that range appear as shown in Figure \ref{incVplot}.  

The region of instability is $k < k_{cr}$, with a maximum growth at $k_m$, where
\begin{mathletters}
\begin{align}
k_{cr} &= 2 \frac { |\dot V_s| \sqrt{-S}}{c_s^2} \\
k_m &= \frac {|\dot V_s| \sqrt{-S}} {c_s^2}.
\end{align}
\end{mathletters}
Compared with Equations \ref{cVks} and Figure \ref{cVplot}, we see that the principal result of removing the effects of compressibility is to eliminate the region of stability near $k = 0$.  We also see that the bending modes now travel asymptotically for high $k$ with the full speed of sound, where previously they moved at $c_s^2 / \sqrt{T}$.  

We note that for $H <<  c_s^2 / |\dot{V}_s|$, the rightmost term in Equation \ref{criticaldisp} becomes very large.  As $H$ becomes very close to zero, one perhaps expects this term to level off at the value in Equation \ref{Vdisp}; we will explore this limit below.

\section{Post-Shock Flow Patterns}

\begin{figure} [tb]
   \includegraphics[width=\columnwidth]{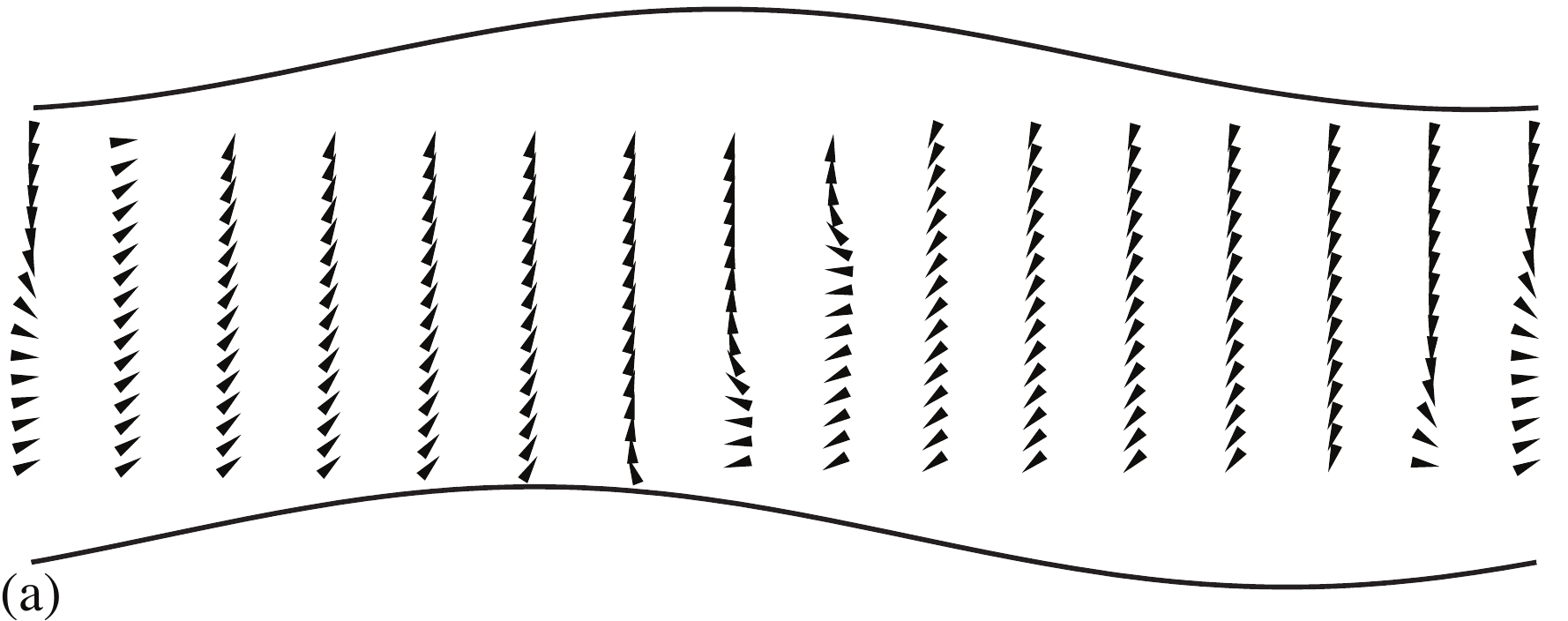} 
   \includegraphics[width=\columnwidth]{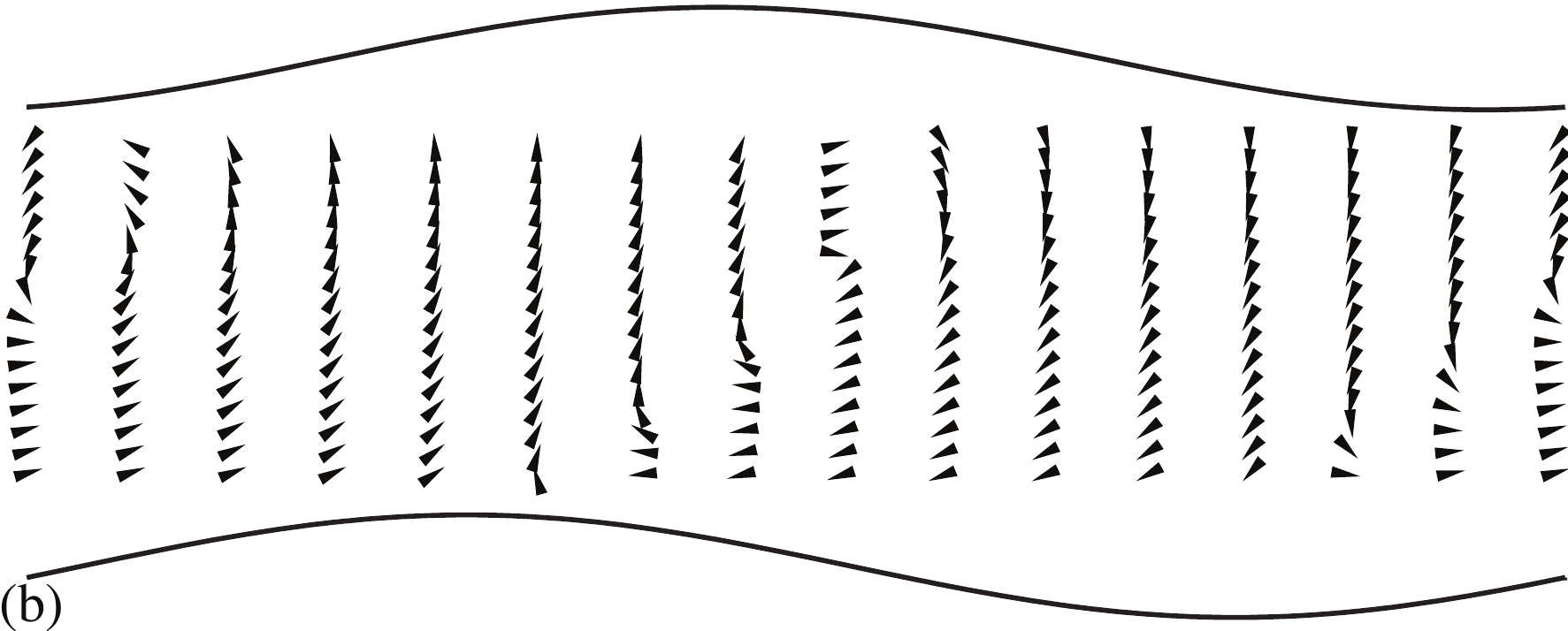} 
   \includegraphics[width=\columnwidth]{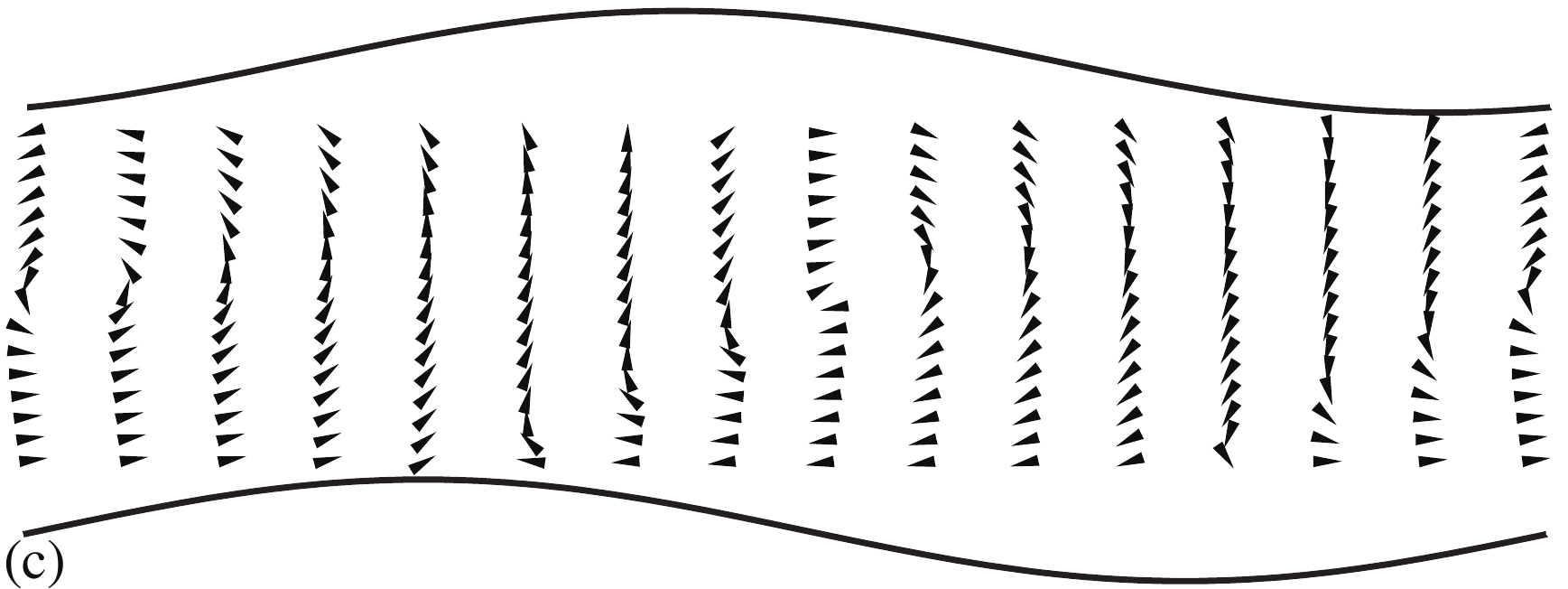} 
   \caption{Numerical solutions of Equation \ref{wavedisp}, showing flow patterns of the perturbation $(u,w)$ within the layer and relative phase of surface perturbations, with (a) $H= 50 \cdot 10^{-6}$ m, (b) $H= 110 \cdot 10^{-6}$ m, (c) $H= 190 \cdot 10^{-6}$ m, for a shock system with $V_s = 120 \cdot 10^3$ m/sec, $\dot{V}_s = - 5 \cdot 10^{12} \textrm{ m/sec}^2$, $\eta = 0.05$, displaying in each case a perturbation with $k = 5210 \textrm{ m}^{-1}$.}
   \label{flows}
\end{figure}

Figure \ref{flows} shows a numerical solution of Equation \ref{wavedisp2} for a shock system with three different thicknesses.  The shock system has a scale height $ c_s^2  / |\dot{V}_s|$ of ${144 \cdot 10^{-6} \textrm{ m}}$.  One can see that for the very thin layer in Fig \ref{flows}(a), the flow pattern is most similar to that of a surface wave.  As the post-shock layer increases in thickness through Figs \ref{flows}(b) and (c), the flow pattern evolves to contain vorticity features.  We speculate that the transition at the scale height corresponds to a layer thickness in which a complete cell is localized.

We remark that in the numerical solution of Equation \ref{wavedisp2} we find that the shock and rear surfaces' perturbations achieve different phase.  Since the fluid inside the layer is constant in density, this will lead to a corresponding perturbation of areal density of the layer that might be observed.  The physical connection is therefore maintained with the theory described by \cite{V83}, in which dynamics causing variation in areal density of the post-shock layer leads to overstability in the shock.  These plots may be compared with Figures 7-10 of \cite{Bertschinger1986}, which show similar vortical structure, though without boundary phase shifting.

\section{Further Considerations and Conclusions}

\subsection{Connections to the Infinitely Thin System}
We have seen that the characteristic fourth-order nature of the Vishniac instability, as derived in Equation \ref{wavedisp}, follows from allowing perturbations on both surfaces of the post-shock layer.
We note that while the Vishniac derivations contain an instability source in the product $\dot{V}_s P_i / \sigma$, our dispersion relation in Equation \ref{wavedisp} contains a source term $\dot{V}_s^2$.  This difference follows from Vishniac's assumption that the post-shock layer is thin and that the difference between thermal backing pressure and ram pressure together with geometric factors (such as spherical divergence of the shock) are the fundamental sources of the deceleration.  We have instead worked with planar shocks and assumed deceleration to stem primarily from mass accumulation and energy loss from the system, for example by strong radiative cooling, and a hydrostatic distribution within the layer to be the dominant contributor to pressure variation.

Despite these differences in approach, we can in fact derive Equation \ref{Vdisp} from Equation \ref{criticaldisp} immediately.  We identify the sound speed at the shock surface with local post-shock fluid variables
\begin{equation} \label{soundspeed}
 c_s^2  = \frac{P(0)}{\rho} = \frac {P_i - \rho \dot{V}_s H}{\rho}
.\end{equation}
We have implicity set the polytropic index $\gamma = 1$, which is consistent with our assumption in Equation \ref{criticaldisp} that we are in the infinitely compressive limit $\eta = 0$.  However, we do not expect Equation \ref{soundspeed} to be in general consistent with our other definitions of $ c_s^2 $, except in the limit of an infinitely thin shell, $H \rightarrow 0$.  Keeping this in mind, we see that inserting Equation \ref{soundspeed} and $\sigma = \rho H$ into the term in square brackets in Equation \ref{criticaldisp}, one obtains Equation \ref{Vdisp}.  Our derivation therefore is found to agree with the earlier results of Vishniac in the appropriate limits.


We comment on the different solutions to Equations \ref{Vdisp}, \ref{cVdisp}, and \ref{criticaldisp}.  The oscillating instability which exists when $\dot V_s < 0$ is the case of interest in which collective modulation of the boundary layers results in the growth of structure.  The non-oscillating instability which appears when $\dot V_s > 0$ is recognized as the Rayleigh-Taylor instability of the rear layer under acceleration.

The non-oscillating solutions of Equations \ref{cVdisp} and \ref{criticaldisp} when $\dot V_s < 0$ but $H >  c_s^2 / |\dot V_s|$ are of a different nature than the other cases.  The system under perturbation was constructed by equating the pressure $P$ immediately behind the shock with the ram pressure of the incoming material.  The pressure profile then decreased hydrostatically with distance from the shock.  When $H$ exceeds a scale height, the most distant pressures obtained in this fashion become negative.  The ``instability'' in this case is a response of the system to inconsistent initial conditions.  
In Vishniac's equation, this corresponds to the case where one assigns $\dot V_s, P_i < 0$.

Compared to Equation \ref{Vdisp}, Equations \ref{wavedisp} and \ref{criticaldisp} have the property of being written in terms of the rear layer height and variables defined locally at the compression front, with few assumptions regarding the structure throughout the layer, while Equation \ref{Vdisp} is properly understood as dealing with quantities averaged over the layer height.  This difference allows one to straightforwardly identify from Equation \ref{criticaldisp} the combination of system variables which lead to the transition at the scale height.  Equation \ref{wavedisp} features the same behavior extended to general post-shock $U$ and finite $\eta$, with appropriate corrections leading to transition at a fraction of the scale height.  We expect the constant density solution to be applicable within a scale height, beyond which modeling the layer as a region of constant density will not be as appropriate as an exponential or self-similar profile.

\subsection{Experimental Observations}

We conclude with some discussion of experiments featuring strongly decelerating planar shocks. Experiments which intend to reproduce this instability must feature sufficient lateral space for the growing perturbations.  Very early in the experiment's evolution, the post-shock layer thickness will be necessarily small, and we assume $H << c_s^2 / |\dot V_s|$.  We see from the results of the proceeding compressible analysis (Equations \ref{cVks} and \ref{cVn}) that to allow maximum growth one must afford the experiment lateral dimensions $\lambda > 2\pi c_s \sqrt{ H / 2 |\dot V_s|}$, where $H$ is a characteristic or average layer thickness of the system.  The evolution will occur within a growth time scale $t = \sqrt{8 H / |\dot V_s|}$.  Conversely, if one wishes to eliminate entirely this instability one should construct an experiment with lateral dimensions $\lambda \lesssim 2.6\  c_s \sqrt{ H / |\dot V_s|}$.  For the experiments discussed above by \cite{Reighard06}, the values of preferred minimum distance and time are approximately 400 - 500 $\mu$m and 9 - 13 ns, conditions which are achievable by the reported experiment.


\acknowledgements

This research was supported by the DOE NNSA under the Predictive Science Academic Alliance Program by grant DE-FC52-08NA28616, the Stewardship Sciences Academic Alliances program by grant DE-FG52-04NA00064, under the National Laser User Facility by grant DE-FG03-00SF22021, and by the Stewardship Science Graduate Fellowship program.

\appendix
\section{Compressible Rear Layer}

We wish to extend the results of Section \ref{SolnsInsideSection} to the investigate the case where the speed of sound varies through the dense layer.  Previously, we assumed a hydrostatic pressure profile on an isothermal layer, which implies the speed of sound varies as 
\begin{equation} \label{nonstantspeedofsound}
c_s^2 = c_s^2(z) = {c_s}_0^2 + \gamma \dot V_s z
\end{equation}
where $\gamma$ is the polytropic index of the layer, and ${c_s}_0^2$ is the speed of sound immediately behind the shock wave.  We revisit the differential equation from \ref{zmom2},
\begin{equation} \label{hosteqn}
(n + U \partial_z)w  =  \partial_z\left( \frac {1}{k^2 + n^2 / c_s^2(z)} (n+U\partial_z)(\partial_z w) \right)
\end{equation}
where we are now treating $c_s^2$ as a function of $z$.

In order to investigate solutions to \ref{hosteqn}, we must first realize that our perturbation ansatz $e^{nt + ikx}$ is no longer valid; either $n$ or $k$ must also vary as a function of $z$.  Since $n$ is our variable of interest, we select $k$ to become $k(z)$.  
We assume that relation between $k$ and $n$ will be linear in $c_s$.
We model the effect by defining
\begin{equation} \label{defs}
\Gamma^2 = k^2(z) c_s(z)^2 + n^2
\end{equation}
where $\Gamma$ is assumed constant.  
We can then rewrite Equation \ref{hosteqn} as
\begin{equation} \label{nonconstdifeq}
\left( \Gamma^2 - \gamma \dot V_s \partial_z - \left({c_s}_0^2  + \gamma \dot V_s z \right) \partial_z^2 \right) \left( n + U \partial_z \right) w = 0.
\end{equation}
We identify the two differential operators
\begin{align}
D_B &= \left( \Gamma^2 - \gamma \dot V_s \partial_z - \left({c_s}_0^2  + \gamma \dot V_s z \right) \partial_z^2 \right) \\
D_t &=  \left( n + U \partial_z \right)
\end{align}
and rewrite Equation \ref{nonconstdifeq} as
\begin{equation} \label{operatoreq}
D_B D_t w = 0.
\end{equation}
It is known from the theory of differential equations that a differential equation in the form above has as its general solution the sum of general solutions of its component operators if they are \textit{permutable}.  The commutator of our operators is nonvanishing, but
\begin{equation} \label{nonconstcommutator}
 [D_B, D_t] =  \gamma \dot V_s U \partial_z^2 
 \end{equation} 
will be neglected, anticipating that we will eventually take the limit of $U$ going to zero.\footnote{
If we do not accept the approximate permutability of $D_B$ and $D_t$, our general solution is found, by use of integrating factors, to be
$$
w = e^{- n z / U} \cdot \frac 1 U \left( \int e^{n z / U} A'\ I_0\!\left(2 \sqrt{ \frac {\Gamma^2 c_s^2(z)}{\gamma^2 \dot V_s^2}}\right) + B'\  K_0\!\left(2 \sqrt{ \frac {\Gamma^2 c_s^2(z)}{\gamma^2 \dot V_s^2}}\right) \ dz \right)
+ C' e^{- n z / U}. 
$$
The arbitrary constants are written with primes to distinguish them from the approximate case.
}

Having eliminated the commutator, one may then consider the sum of general solutions of each independent operator as the complete general solution to the combined equation.  The general solution for $D_t$ is $C e^{- n z / U}$.  The general solution of $D_B$ can be found by a change of variables to
$$\zeta = 2 \sqrt{ \frac {\Gamma^2 {c_s}^2(z)}{\gamma^2 \dot V_s^2}}$$
to cast $D_B$ as
\begin{equation}
D_B = -\Gamma^2 \left( \partial_\zeta^2 + \frac {1} {\zeta} \partial_\zeta - 1 \right)
\end{equation}
which is the operator corresponding to the modified Bessel equation.  Solutions to $D_B$ are of the form $A\ I_0\!\left(2 \sqrt{ \frac {\Gamma^2 c_s^2(z)}{\gamma^2 \dot V_s^2}}\right) + B\ K_0\!\left(2 \sqrt{ \frac {\Gamma^2 c_s^2(z)}{\gamma^2 \dot V_s^2}}\right)$, where $I_0$ and $K_0$ are the modified Bessel functions.

We take as our approximate general solution
\begin{equation} \label{nonconstgeneralsoln}
w = A\ I_0\!\left(2 \sqrt{ \frac {\Gamma^2 c_s^2(z)}{\gamma^2 \dot V_s^2}}\right) + B\ K_0\!\left(2 \sqrt{ \frac {\Gamma^2 c_s^2(z)}{\gamma^2 \dot V_s^2}}\right) + C e^{- n z / U}.
\end{equation}
The new basis for $w$ written with modified Bessel's functions is less dissimilar to the previous basis, Equation \ref{solns}, than it appears at first glance.  To first order in $\delta$s and zeroth order in $U/c_s$, the differential forms of the boundary conditions are changed only cosmetically.
\begin{mathletters}
\begin{align}
0 &=  \left. \left( n^2c_s^2(z) \partial_z - \Gamma^2 \dot V_s \right) w\ \right|_{z=H} \label{nonconstrearcond} \\
0 &=  \left.\left( \frac {U c_s^2(z)}  {\Gamma^2} \partial^2_z + \frac {n c_s^2(z) }{ \Gamma^2} \partial_z  -  \left( \frac {\dot{V}_s }{ n (1 - \eta)} + 2 U\right)\right)  w\ \right|_{z=0} \label{nonconstprescond} \\
0 &= \left. \left( \frac {c_s^2(z) }{ \Gamma^2} \partial_z - \frac {V_s }{n(1-\eta)} \right) w   \frac{{}}{}\right|_{z=0} \label{nonconstshockcond}
\end{align}
\end{mathletters}

%

To obtain the dispersion relation from the boundary conditions, we note that
$$
j(H) =  {j(0)} \frac{ c_{s0}}{c_s(H)}= {j(0)} {\sqrt{1 - \frac { \gamma \dot V_s H }{ c_s^2(H)}}}.
$$
and evaluate $w$ to obtain
\begin{equation} \label{nonconstmatrix}
\left|
\begin{array}{ccc}
n^2 K_1(\zeta_H) - j \dot{V}_s K_0(\zeta_H)\frac { {c_s}_0}{c_s(H)} & -n^2 I_1(\zeta_H)  - j \dot{V}_s I_0(\zeta_H) \frac { {c_s}_0}{c_s(H)} & 0 \\
\parbox{1.2in}{$ -\frac{n }{j}\left(1 - \gamma \frac{ U \dot V_s}{n {c_s^2}_0}  \right) K_1 (\zeta_0) \\ + \left(U +\frac{{\dot V_s} }{n (1 - \eta)} \right) K_0(\zeta_0)$} 
 &
 \parbox{1.2in}{$ \frac{n }{j}\left( 1 - \gamma \frac{ U \dot V_s}{n {c_s^2}_0} \right) I_1(\zeta_0) \\ + \left( U  +\frac{\dot V_s}{n(1 - \eta)} \right) I_0(\zeta_0)$} 
 & 2U + \frac{\dot V_s}{n(1 - \eta)} \\
 \frac {1} {j} K_1(\zeta_0) - \frac{V_s}{n( 1 - \eta)} K_0(\zeta_0) & - \frac {1} {j} I_1(\zeta_0) - \frac{V_s}{n(1- \eta)} I_0(\zeta_0) & -\frac{n}{ U j^2 }-\frac{ V_s}{n(1-\eta)}
 \end{array}
\right| = 0 .
\end{equation}
In Equation \ref{nonconstmatrix}, $j = j(0)$, $\zeta_H = 2 \sqrt{ \frac {\Gamma^2 c_s^2(H)}{\gamma^2 \dot V_s^2}}$, and $\zeta_0 = 2 \sqrt{ \frac {\Gamma^2 {c_s}_0^2 }{\gamma^2 \dot V_s^2}}$.

Equation \ref{nonconstmatrix} reduces to the dispersion relation with constant speed of sound in the limit of $\gamma \rightarrow 0$. We see that the only substantial changes are in the terms incorporating the effect of layer height $H$ and the appearance of two terms of $\gamma \frac{ U \dot V_s}{n {c_s^2}}$.  The latter of these is the same term which was neglected previously in writing Equation \ref{nonconstcommutator}, and will be neglected here for consistency.


In analogy to Section \ref{ThinLayerSection}, we take the limit of $U \rightarrow 0$, $V_s \rightarrow \infty$, $U V_s \rightarrow c_{s0}^2$, and write the dispersion relation as
\begin{mathletters}
\begin{equation}
0 = (1 - \eta) n^2 + j^2 c_{s0}^2 + j \dot V_s \left( \frac{n^2 - j \dot V_s F_1}{n^2 F_2 - j \dot V_s F_3} \right)
\end{equation}
where
\begin{align}
F_1 &= \frac {I_0(\zeta_0) K_0(\zeta_H)- K_0(\zeta_0) I_0(\zeta_H)}{I_0(\zeta_0) K_1(\zeta_H) + K_0(\zeta_0) I_1(\zeta_H)} \frac{{c_s}_0}{c_s(H)} \\
F_2 & = \frac {I_1(\zeta_0) K_1(\zeta_H) - K_1(\zeta_0) I_1(\zeta_H)}{I_0(\zeta_0) K_1(\zeta_H) + K_0(\zeta_0)I_1(\zeta_H)}\\
F_3 &= \frac {I_1(\zeta_0) K_0(\zeta_H) + K_1(\zeta_0) I_0(\zeta_H)}{I_0(\zeta_0) K_1(\zeta_H) + K_0(\zeta_0) I_1(\zeta_H)} \frac {{c_s}_0 }{c_s(H)}
\end{align}
\end{mathletters}
Compared to the previous dispersion relation in Equation \ref{wavedisp}, $F_1$ and $F_2$ are analogous to $\tanh(jH)$ and $F_3$ was previously equal to one.  These identities are preserved if we assign the cylinder functions $I_{0, 1}(\zeta_0) = 1, I_{0,1}(\zeta_H) = e^{- j H}$,  $K_{0, 1}(\zeta_0) = 1, K_{0,1}(\zeta_H) = e^{ j H}$, and $\gamma=0$ (and therefore $c_s(H) = c_{s0}$).

The effects of the changing speed of sound can be approximately included in, for example, Equation \ref{criticaldisp}, by writing
\begin{eqnarray} \label{amplifieddisp}
n^4 + n^2 k^2  {c_s}_0^2 -  \frac {{c_s}_0 }{c_s(H)}  k^2 \dot{V}_s^2\left[1 + \frac { c_s^2 / \dot{V}_s} {H} \right] = 0
\end{eqnarray}
where $k = k|_{z=0}$.  The diminishing speed of sound with rising layer height evidently amplifies the instability.  Physically, this comes from the fact that for a given $n$, the wavelength of  sound waves will be shorter in the region of lower sound speed.  This leads to an increased $k$ on the rear, instability-forming boundary condition.



\begin{thebibliography}{14}
\expandafter\ifx\csname natexlab\endcsname\relax\def\natexlab#1{#1}\fi
\expandafter\ifx\csname bibnamefont\endcsname\relax
  \def\bibnamefont#1{#1}\fi
\expandafter\ifx\csname bibfnamefont\endcsname\relax
  \def\bibfnamefont#1{#1}\fi
\expandafter\ifx\csname citenamefont\endcsname\relax
  \def\citenamefont#1{#1}\fi
\expandafter\ifx\csname url\endcsname\relax
  \def\url#1{\texttt{#1}}\fi
\expandafter\ifx\csname urlprefix\endcsname\relax\def\urlprefix{URL }\fi
\providecommand{\bibinfo}[2]{#2}
\providecommand{\eprint}[2][]{\url{#2}}

\bibitem[{\citenamefont{{Vishniac}}(1983)}]{V83}
\bibinfo{author}{\bibfnamefont{E.~T.} \bibnamefont{{Vishniac}}},
  \bibinfo{journal}{\apj} \textbf{\bibinfo{volume}{274}}, \bibinfo{pages}{152}
  (\bibinfo{year}{1983}).

\bibitem[{\citenamefont{{Vishniac} and {Ryu}}(1989)}]{V89}
\bibinfo{author}{\bibfnamefont{E.~T.} \bibnamefont{{Vishniac}}}
  \bibnamefont{and} \bibinfo{author}{\bibfnamefont{D.}~\bibnamefont{{Ryu}}},
  \bibinfo{journal}{\apj} \textbf{\bibinfo{volume}{337}}, \bibinfo{pages}{917}
  (\bibinfo{year}{1989}).

\bibitem[{\citenamefont{{Bertschinger}}(1986)}]{Bertschinger1986}
\bibinfo{author}{\bibfnamefont{E.}~\bibnamefont{{Bertschinger}}},
  \bibinfo{journal}{\apj} \textbf{\bibinfo{volume}{304}}, \bibinfo{pages}{154 }
  (\bibinfo{year}{1986}).

\bibitem[{\citenamefont{{Kushnir} et~al.}(2005)\citenamefont{{Kushnir},
  {Waxman}, and {Shvarts}}}]{KWS05}
\bibinfo{author}{\bibfnamefont{D.}~\bibnamefont{{Kushnir}}},
  \bibinfo{author}{\bibfnamefont{E.}~\bibnamefont{{Waxman}}}, \bibnamefont{and}
  \bibinfo{author}{\bibfnamefont{D.}~\bibnamefont{{Shvarts}}},
  \bibinfo{journal}{\apj} \textbf{\bibinfo{volume}{634}}, \bibinfo{pages}{407}
  (\bibinfo{year}{2005}).

\bibitem[{\citenamefont{Drake}(2006)}]{DrakeHEDP}
\bibinfo{author}{\bibfnamefont{R.~P.}~\bibnamefont{Drake}},
  \emph{\bibinfo{title}{High-Energy-Density Physics}} (\bibinfo{publisher}{Springer}, \bibinfo{year}{2006}).


\bibitem[{\citenamefont{Reighard et~al.}(2006)\citenamefont{Reighard, Drake,
  Dannenberg, Kremer, Grosskopf, Harding, Leibrandt, Glendinning, Perry,
  Remington et~al.}}]{Reighard06}
\bibinfo{author}{\bibfnamefont{A.~B.} \bibnamefont{Reighard}},
  \bibinfo{author}{\bibfnamefont{R.~P.} \bibnamefont{Drake}},
  \bibinfo{author}{\bibfnamefont{K.~K.} \bibnamefont{Dannenberg}},
  \bibinfo{author}{\bibfnamefont{D.~J.} \bibnamefont{Kremer}},
  \bibinfo{author}{\bibfnamefont{M.}~\bibnamefont{Grosskopf}},
  \bibinfo{author}{\bibfnamefont{E.~C.} \bibnamefont{Harding}},
  \bibinfo{author}{\bibfnamefont{D.~R.} \bibnamefont{Leibrandt}},
  \bibinfo{author}{\bibfnamefont{S.~G.} \bibnamefont{Glendinning}},
  \bibinfo{author}{\bibfnamefont{T.~S.} \bibnamefont{Perry}},
  \bibinfo{author}{\bibfnamefont{B.~A.} \bibnamefont{Remington}},
  \bibnamefont{et~al.}, \bibinfo{journal}{Phys. Plasmas}
  \textbf{\bibinfo{volume}{13}}, \bibinfo{pages}{082901}
  (\bibinfo{year}{2006}).

\bibitem[{\citenamefont{Bouquet et~al.}(2004)\citenamefont{Bouquet, St\'ehl\'e,
  Koenig, Chi\`eze, Benuzzi-Mounaix, Batani, Leygnac, Fleury, Merdji, Michaut
  et~al.}}]{BouquetPRL}
\bibinfo{author}{\bibfnamefont{S.}~\bibnamefont{Bouquet}},
  \bibinfo{author}{\bibfnamefont{C.}~\bibnamefont{St\'ehl\'e}},
  \bibinfo{author}{\bibfnamefont{M.}~\bibnamefont{Koenig}},
  \bibinfo{author}{\bibfnamefont{J.-P.} \bibnamefont{Chi\`eze}},
  \bibinfo{author}{\bibfnamefont{A.}~\bibnamefont{Benuzzi-Mounaix}},
  \bibinfo{author}{\bibfnamefont{D.}~\bibnamefont{Batani}},
  \bibinfo{author}{\bibfnamefont{S.}~\bibnamefont{Leygnac}},
  \bibinfo{author}{\bibfnamefont{X.}~\bibnamefont{Fleury}},
  \bibinfo{author}{\bibfnamefont{H.}~\bibnamefont{Merdji}},
  \bibinfo{author}{\bibfnamefont{C.}~\bibnamefont{Michaut}},
  \bibnamefont{et~al.}, \bibinfo{journal}{Phys. Rev. Lett.}
  \textbf{\bibinfo{volume}{92}}, \bibinfo{pages}{225001}
  (\bibinfo{year}{2004}).

\bibitem[{\citenamefont{Bozier et~al.}(1986)\citenamefont{Bozier, Thiell,
  Le~Breton, Azra, Decroisette, and Schirmann}}]{BozierPRL}
\bibinfo{author}{\bibfnamefont{J.~C.} \bibnamefont{Bozier}},
  \bibinfo{author}{\bibfnamefont{G.}~\bibnamefont{Thiell}},
  \bibinfo{author}{\bibfnamefont{J.~P.} \bibnamefont{Le~Breton}},
  \bibinfo{author}{\bibfnamefont{S.}~\bibnamefont{Azra}},
  \bibinfo{author}{\bibfnamefont{M.}~\bibnamefont{Decroisette}},
  \bibnamefont{and}
  \bibinfo{author}{\bibfnamefont{D.}~\bibnamefont{Schirmann}},
  \bibinfo{journal}{Phys. Rev. Lett.} \textbf{\bibinfo{volume}{57}},
  \bibinfo{pages}{1304} (\bibinfo{year}{1986}).

\bibitem[{\citenamefont{Remington et~al.}(2006)\citenamefont{Remington, Drake,
  and Ryutov}}]{RemingtonRMP}
\bibinfo{author}{\bibfnamefont{B.~A.} \bibnamefont{Remington}},
  \bibinfo{author}{\bibfnamefont{R.~P.} \bibnamefont{Drake}}, \bibnamefont{and}
  \bibinfo{author}{\bibfnamefont{D.~D.} \bibnamefont{Ryutov}},
  \bibinfo{journal}{Reviews of Modern Physics} \textbf{\bibinfo{volume}{78}},
  \bibinfo{eid}{755} 
   (\bibinfo{year}{2006}).

\bibitem[{\citenamefont{Hayes and Probstein}(1966)}]{Hayes}
\bibinfo{author}{\bibfnamefont{W.}~\bibnamefont{Hayes}} \bibnamefont{and}
  \bibinfo{author}{\bibfnamefont{R.}~\bibnamefont{Probstein}},
  \emph{\bibinfo{title}{Hypersonic Flow Theory}} (\bibinfo{publisher}{Academic
  Press}, \bibinfo{year}{1966}).


\bibitem[{\citenamefont{{Liang and Keilty}}(2000)}]{Liang}
\bibinfo{author}{\bibfnamefont{Edison~Liang} \bibnamefont{and} \bibnamefont{{Katherine~Keilty}}},
  \bibinfo{journal}{\apj}
  \textbf{\bibinfo{volume}{533}}, \bibinfo{pages}{890} (\bibinfo{year}{2000}).


\bibitem[{\citenamefont{{Freeman}}(1955)}]{Freeman1955}
\bibinfo{author}{\bibfnamefont{N.~C.} \bibnamefont{{Freeman}}},
  \bibinfo{journal}{Royal Society of London Proceedings Series A}
  \textbf{\bibinfo{volume}{228}}, \bibinfo{pages}{341} (\bibinfo{year}{1955}).


\bibitem[{\citenamefont{{Vishniac}}(1995)}]{V1995}
\bibinfo{author}{\bibfnamefont{E.~T.} \bibnamefont{{Vishniac}}},
  \bibinfo{journal}{New York Academy Sciences Annals}
  \textbf{\bibinfo{volume}{773}}, \bibinfo{pages}{70} (\bibinfo{year}{1995}).

\bibitem[{\citenamefont{{Ishizaki and Nishimura}}(1997)}]{IshizakiPRL}
\bibinfo{author}{\bibfnamefont{R.} \bibnamefont{{Ishizaki}}}
\bibnamefont{and}
  \bibinfo{author}{\bibfnamefont{K.}~\bibnamefont{Nishimura}},
  \bibinfo{journal}{\prl}
  \textbf{\bibinfo{volume}{78}}, \bibinfo{pages}{1920} (\bibinfo{year}{1997}).


\end{thebibliography}
\end{document}